\documentclass[12pt,preprint]{aastex}

\shorttitle{Evolution of the Obscuration in AGN}
\shortauthors{Treister \& Urry}

\begin{document}

\title{The Evolution of Obscuration in Active Galactic Nuclei}

\author{Ezequiel Treister\altaffilmark{1,2} and C. Megan Urry\altaffilmark{3,4}}

\altaffiltext{1}{European Southern Observatory, Casilla 19001, Santiago 19, Chile. E-mail: etreiste@eso.org}
\altaffiltext{2}{Departamento de Astronom\'{\i}a, Universidad de Chile, Casilla 36-D, Santiago, Chile.}
\altaffiltext{3}{Yale Center for Astronomy \& Astrophysics, Yale University,
P.O. Box 208121, New Haven, CT 06520}
\altaffiltext{4}{Department of Physics, Yale University, P.O. Box 208120, New Haven, CT 06520.}


\begin{abstract}
In order to study the evolution of the relative fraction of obscured
Active Galactic Nuclei (AGN) we constructed the largest sample to date
of AGN selected in hard X-rays. The full sample contains 2341
X-ray-selected AGN, roughly 4 times the largest previous samples
studied in this connection. Of these, 1229 (53\%) have optical
counterparts for which redshifts are available; these span the
redshift range $z$=0-4. The observed fraction of obscured AGN declines
only slightly with redshift. Correcting for selection bias, we find
that the intrinsic fraction of obscured AGN must actually increase
with redshift, as (1+$z$)$^{\alpha}$, with
$\alpha$$\simeq$0.4$\pm$0.1. This evolution is consistent with the
integrated X-ray background, which provides the strongest constraints
at relatively low redshift, $z\sim$1. Summing over all AGN, we
estimate the bolometric AGN light to be 3.8 nW~m$^{-2}$~sr$^{-1}$, or
$\lesssim$8\% of the total extragalactic light. 
Together with the observed black hole mass density in the local
Universe, this implies an accretion efficiency of $\eta \sim$0.1-0.2,
consistent with the values typically assumed.

\end{abstract}


\section{Introduction}

It is now clear from the deepest Chandra and XMM observations (e.g.,
\citealp{giacconi01,brandt01}) that a combination of obscured and
unobscured Active Galactic Nuclei (AGN) are needed to explain the
observed properties of the extragalactic X-ray background (XRB). While
unobscured AGN are readily detected using, for example, optical
color-selection techniques, the optical-UV-soft X-ray signatures of
nuclear activity are not visible in obscured AGN, making their
detection and identification much harder. The effects of obscuration
are less important at high energies, hence deep, hard X-ray
(2--10~keV) surveys with Chandra and XMM have revealed a less biased
view of the AGN population (e.g, \citealp{barger03} and references
therein), although they are still insensitive to the most obscured
sources, like Compton-thick AGN (e.g., \citealp{worsley05}). However,
these surveys rely mostly on optical spectroscopy to find redshifts
and thus luminosities, hence they preferentially exclude obscured AGN,
which have fainter optical counterparts (\citealp{treister04} and
references therein). This means the observed spectroscopically
identified population is not representative of the underlying AGN
population.

One fundamental ingredient in our understanding of the AGN population
is the ratio of obscured to unobscured AGN, and whether this ratio
depends on parameters like intrinsic luminosity or
redshift. According to the unification paradigm
\citep{antonucci93,urry95}, the ratio depends on the geometry of the
circumnuclear obscuring material. In the simplest version of
unification, this ratio is independent of source properties like
luminosity or redshift. However, there are physical reasons to expect
a dependence; for example, more luminous accretion disk emission may
correspond to a larger dust sublimation radius
\citep{lawrence91,simpson05}, thus increasing the fraction of
unobscured AGN at higher luminosities. Similarly, if the dust content
of the AGN host galaxy is important for the obscuration of the central
engine, a dependence of the obscured AGN fraction on redshift would be
expected (e.g., \citealp{ballantyne06} and references therein), given
the evolution of the amount of dust in galaxies. Thus, a measurement
of the fraction of obscured AGN and its possible dependence on
critical parameters can be used to study AGN structure and to probe
the connection between AGN activity and the formation of the host
galaxy.

The XRB also constrains the fraction of obscured AGN. Early AGN
population synthesis models for the XRB (e.g.,
\citealp{comastri95,gilli99}) required a strong increase of the
fraction with redshift, while recent versions (e.g.,
\citealp{ueda03,treister05b}) assume only a decrease with increasing
luminosity, although \citet{ballantyne06} concluded that a evolving
obscured AGN fraction provides a better fit.  Deep X-ray surveys
confirm that the obscured AGN fraction depends on luminosity
\citep{ueda03,steffen03}, while the dependence on redshift
is less clear. \citet{lafranca05}, using a larger heterogeneous sample
and X-ray spectral fitting to classify AGN, concluded that the
relative fraction of obscured AGN increases with redshift. However,
although \citet{akylas06} found a similar evolution, they concluded
from simulations that such a correlation can be induced by the K
correction of the X-ray spectra.

In this work, we generate an AGN sample with high optical
spectroscopic completeness, the largest such sample to date by a
factor of $\sim$4, in order to study the evolution of the obscured AGN
fraction. Using optical spectroscopy we characterize AGN as
unobscured if they have broad emission lines, and we measure the
dependence of the obscured fraction on redshift.  Throughout this
paper we assume $H_0=70$ km s $^{-1}$ Mpc$^{-1}$, $\Omega_m=0.3$ and
$\Omega_\Lambda =0.7$.

\section{The Sample}

\begin{figure}
\figurenum{1}
\plottwo{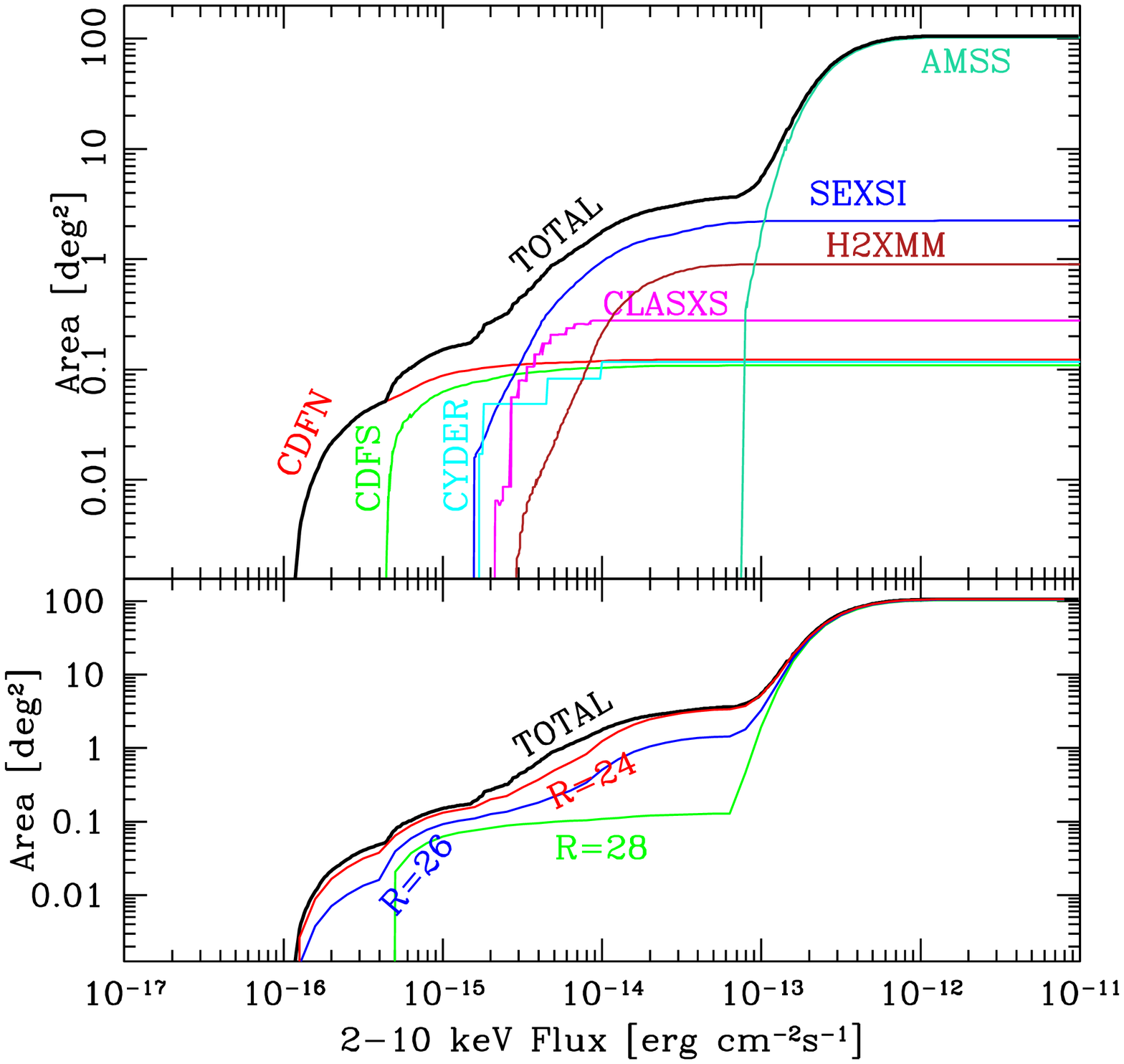}{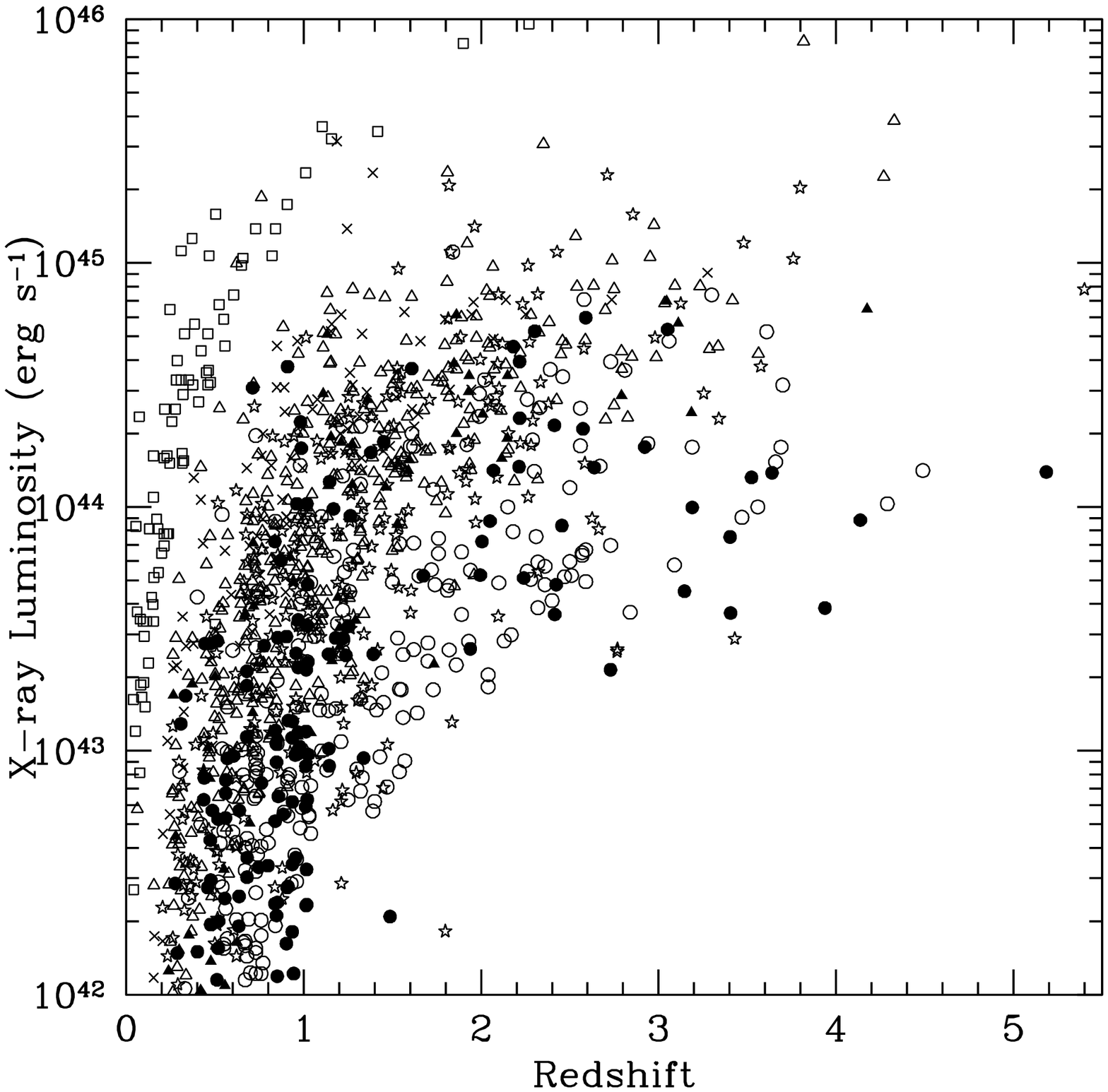}
\caption{{\it Left figure.} {\it Upper panel}: Area versus hard X-ray flux relation for
the surveys used in this work ({\it light lines}) and for the total
sample ({\it thick line}). {\it Lower panel}: Total effective area as
a function of X-ray flux and optical magnitude, taking into account
the spectroscopic incompleteness of each survey (see text for
details), for $R$=24, 26 and 28.  These curves are used to compute the
expected fraction of identified obscured AGN for an intrinsically
non-evolving ratio.{\it Right figure}. X-ray luminosity versus
redshift relation for the sources in our sample from the CDF-N ({\it
filled circles}), CDF-S ({\it open circles}), SEXSI ({\it open
triangles}), CYDER ({\it filled triangles}), CLASXS ({\it stars}),
HELLAS2XMM ({\it crosses}) and AMSS ({\it squares}) surveys.}
\label{area}
\end{figure}

To distinguish between the effects of redshift and luminosity, our
sample needs to probe a range of luminosities at each redshift over a
reasonable a range of redshifts. Wide area, shallow X-ray surveys
sample moderate luminosity AGN at low redshifts and only high
luminosity sources up to high redshifts, while deep pencil-beam
surveys are useful to study the moderate luminosity population at high
redshifts, but only provide a small number of the more rare,
high-luminosity sources. Combining the two covers the
luminosity-redshift plane effectively (see Fig.~\ref{area}). Here we
combine seven wide and deep surveys, for a total sample of 2341 AGN
selected in the hard X-ray band. (We define an AGN as an X-ray source
more luminous than $L_{2-10~\rm{keV}}=10^{42}$~ergs/s.) Table 1
summarizes the surveys used in this work and their main
characteristics. When necessary, hard X-ray fluxes in the 2--8~keV
band were converted to the 2--10~keV range assuming a power-law
spectrum of the form dN/dE$\propto$E$^{-\Gamma}$ with $\Gamma$=1.7,
consistent with the average observed AGN spectrum (e.g.,
\citealp{nandra97}). The total area of this super-sample as a function
of X-ray flux is shown in Fig.~\ref{area}.

To classify the sample by optical spectroscopy requires high
spectroscopic completeness; all the constituent surveys used here are
at least 40\% complete (most are much higher) and a total of 1229
sources (53\% of our sample) have reliable redshifts.  These are
spectroscopic redshifts except in the Chandra Deep Field South, where
the photometric classification by \citet{zheng04} was used when
spectroscopic information was not available (i.e., for about half the
227 objects; the photometric redshifts include Combo-17 data and so
are quite accurate). Sources were classified as unobscured AGN when
broad emission lines were present in the optical spectrum (or using
the optical/IR continuum shape for the CDF-S AGN with photometry
only); the 631 sources (51\%) without broad emission lines and with
X-ray luminosities greater than 10$^{42}$~erg~s$^{-1}$ were classified
as obscured AGN.  This is a more robust classification scheme than
using X-ray-determined $N_H$ because there are no K correction
effects, since lines are present in the optical spectrum at most
redshifts (line dilution by the host galaxy is not important;
\citep{barger05}.

For obvious reasons, the spectroscopic completeness of any survey
depends on the brightness of the optical counterparts. To describe
this effect quantitatively, we characterize the identified fraction of
each constituent survey by a simple 3-parameter function that is
constant at bright magnitudes and declines linearly to faint
magnitudes. These 3 independent parameters (break and limiting
magnitudes in the R band, and maximum completeness at bright fluxes)
provide a very good description of the identified fraction in all 7
surveys, with values for the reduced $\chi^2$ lower than 0.5 in each
case. The fitted parameters for each survey are given in Table 1. The
effective area as a function of X-ray flux and optical magnitude was
then calculated by weighting the area versus X-ray flux relation for
each survey by its sensitivity (i.e., the fraction of identified AGN)
at each optical magnitude, and summing the results. The total
effective area for the super-sample is shown in Fig.~\ref{area}.

With the selection function of the super-sample quantified, we can now
interpret the observed demographics of X-ray sources. The observed
fraction of obscured AGN as a function of redshift is shown (data
points) in the upper panel of Fig.~\ref{red_ratio}. This fraction
remains almost constant at a value of $\sim$0.6-0.7 up to $z$=1.5 and
drops to $\sim$0.2-0.3 at higher redshifts. A much steeper decline is
expected because the measured redshifts require an optical spectrum.
For obscured AGN the optical emission is dominated by the host galaxy,
which can be studied spectroscopically up to $z\sim$1 but then becomes
too faint for even 8m-class telescopes, hence the decline at high
redshift \citep{treister04}.

To account quantitatively for this selection effect, we
calculate the ratio one should observe for an intrinsically
non-evolving population, taking into account the effects of the
sensitivity and completeness of each survey. This was done using the
AGN population synthesis model of \citet{treister04} as modified by
\citet{treister05b}. This model explains at the same time the X-ray,
optical and infrared number counts of AGN in the Great Observatories
Origins Deep Survey \citep{treister04,treister06} and the spectral
shape and intensity of the extragalactic XRB \citep{treister05b}. In
practice, most of the model parameters are irrelevant to the present
calculation, as it depends mainly on the host galaxy luminosity and
evolution. We assumed an $L_*$ luminosity and a Sc-like evolution as
given by \citet{poggianti97}. (We checked that the results do not
depend significantly on the type of host-galaxy evolution assumed.)
We also assume a local ratio of obscured to unobscured AGN of
$\sim$3:1, in agreement with observations (e.g.,
\citealp{risaliti99}), and this ratio was assumed to decline with
increasing luminosity but to remain constant with redshift. The
expected fraction of obscured AGN as a function of redshift for the
observational parameters of our super-sample declines sharply above
$z\sim1$ (line, upper panel, Fig.~\ref{red_ratio}), mainly because of
spectroscopic incompleteness.

\section{Results and Discussion}

\begin{figure}
\figurenum{2}
\plotone{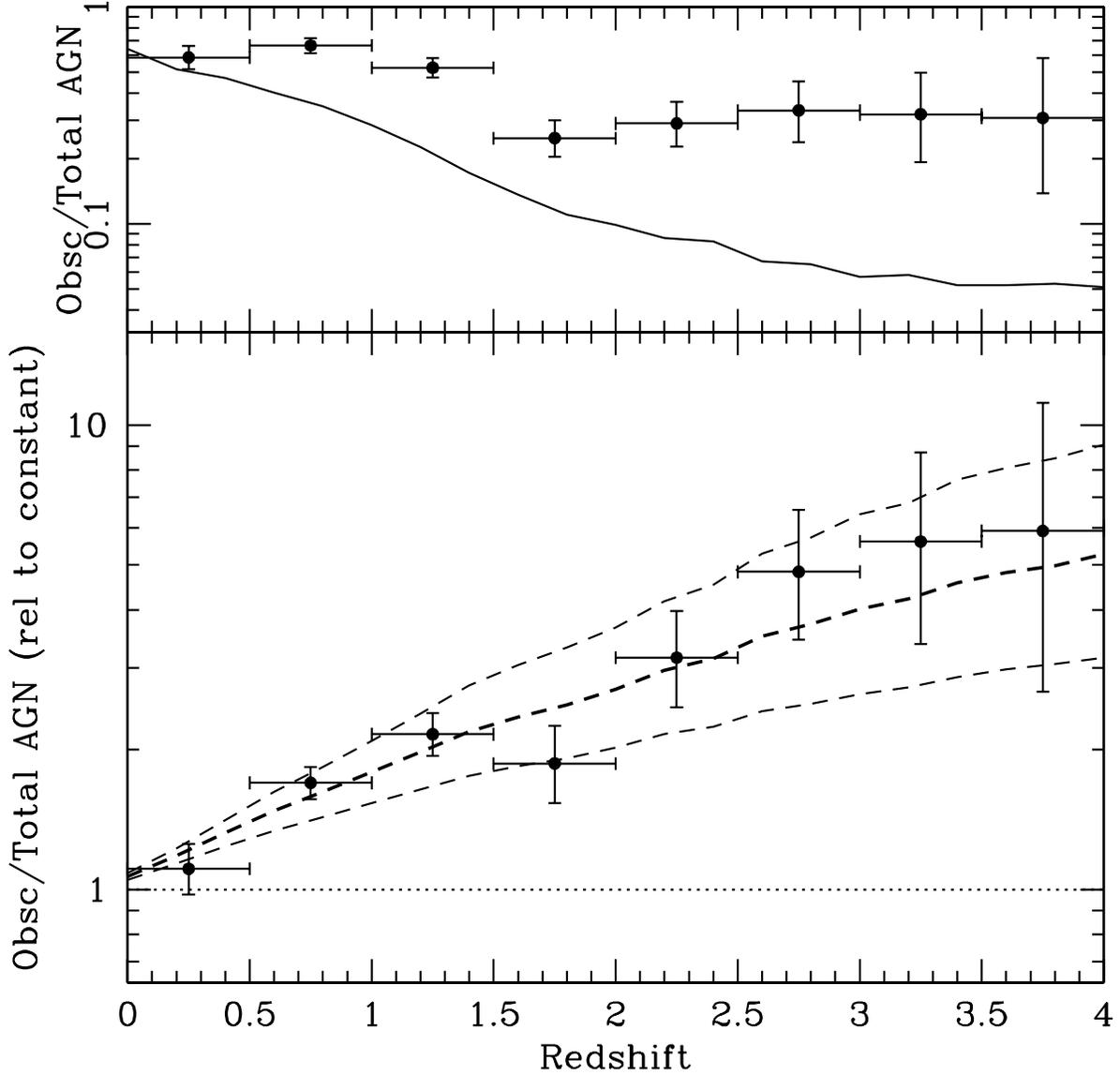}
\caption{{\it Upper panel}: Observed fraction of obscured AGN as a function of
redshift. {\it Data points}: direct measurements from our super-sample
of 1229 optically-classified AGN. {\it Solid line}: Expected fraction
as a function of redshift for an intrinsically non-evolving ratio,
taking into account the effects of spectroscopic incompleteness and
effective area as a function of X-ray flux.  {\it Lower panel}:
Fraction of obscured AGN relative to the expectations for a
non-evolving obscured AGN ratio, incorporating the effects of
spectroscopic incompleteness. A significant increase with redshift is
clearly seen. For an intrinsic evolution of the form
(1+$z$)$^{\alpha}$, the {\it thick dashed line} shows $\alpha$=0.4 and
the {\it thin dashed lines} show $\alpha$=0.5 (upper) and 0.3
(lower). }
\label{red_ratio}
\end{figure}

The observed fraction of obscured AGN at high redshift is higher than
expected if the intrinsic fraction does not evolve.  In the bottom
panel of Fig.~\ref{red_ratio} we show the obscured AGN fraction
relative to the expected value.  Clearly, it increases significantly
with redshift, roughly as (1+$z$)$^\alpha$, with $\alpha$=0.3-0.5
(thin dashed lines, bottom panel, Fig.~\ref{red_ratio}; best fit,
$\alpha$$\simeq$0.4, thick dashed line).  This value of $\alpha$ does
not change significantly if a different host galaxy evolution is
assumed, and it is consistent with the value of 0.3 reported by
\citet{ballantyne06}. Similar evolution was found by \citet{lafranca05}
and \citet{akylas06}, based on a much smaller sample, but we classify
AGN via optical spectroscopy rather than X-ray spectral fitting and so
avoid their K-correction bias.

The XRB also constrains the relative fraction of obscured AGN and its
evolution. The strongest constraints come from 1--10~keV, since at
high energies the effects of obscuration are less important, and
fortunately, the spectrum of the XRB in this energy range has been
well measured. Based on XMM observations, \citet{deluca04} reported an
integrated XRB flux in the 2--10~keV band of
2.24$\pm$0.16$\times$10$^{-11}$~erg~cm$^{-2}$~s$^{-1}$. Integrating
our AGN population synthesis model gives, for $\alpha$=0 (i.e., a
non-evolving obscured AGN fraction),
2.3$\times$10$^{-11}$~erg~cm$^{-2}$~s$^{-1}$. Incorporating the
evolution with redshift, we obtain 2.1, 2.04, and 2.00
$\times$10$^{-11}$~erg~cm$^{-2}$~s$^{-1}$ for $\alpha$=0.3, 0.4, and
0.5, respectively. (Larger values of $\alpha$ imply a lower integrated
flux at low energies because of a higher relative fraction of obscured
AGN, in which most of the soft X-ray emission is absorbed.)

Similarly, using $Chandra$ observations, \citet{hickox06} report an
integrated flux value of
1.7$\pm$0.2$\times$10$^{-11}$~erg~cm$^{-2}$~s$^{-1}$ for the 2--8~keV
band. In comparison, for $\alpha$=0 we find
1.8$\times$10$^{-11}$~erg~cm$^{-2}$~s$^{-1}$ in the 2--8~keV band, and
1.7, 1.6, and $1.5 \times$10$^{-11}$~erg~cm$^{-2}$~s$^{-1}$ for
$\alpha$=0.3, 0.4, and 0.5, respectively. We conclude that the XRB
favors $\alpha$$\simeq$0.3, in good agreement with the value found
from the observed fraction.

Since forming galaxies may be expected to have more dust,
the increase in the relative fraction of obscured AGN at high redshift may
be due to an increase in the contribution to obscuration by galactic
dust. Combining hard X-ray and mid-infrared observations,
\citet{lutz04} found a similar ratio of hard X-ray to mid-infrared
flux for obscured and unobscured AGN, contrary to the predictions of
the simplest AGN unification paradigm, in which the obscuration comes
from the dusty torus and therefore the mid-infrared emission is
reduced due to self-absorption. This result can be explained if the
obscuration comes from a much more extended region, i.e., kiloparsec,
galactic scales rather than a compact parsec-scale torus. Furthermore,
signatures for extended absorbing regions have been detected in nearby
galaxies like NGC 1068 (e.g., \citealp{bock98}) and NGC 4151
\citep{radomski03}. Heavy absorption at kpc-scales has routinely been
found in ultra-luminous infrared galaxies (ULIRGs), which suffer a
very strong evolution (e.g., \citealp{saunders90}). Hence, it seems
likely that the change in the relative fraction of obscured AGN could
be related to galactic-scale absorption, in particular since some
ULIRGs also contain an obscured AGN (e.g., Arp 220;
\citealp{iwasawa05}).

The contribution of AGN to the bolometric energy budget of the
Universe is a matter of debate. While it is now clear that AGN are the
major contributor in X-rays, they only constitute $\sim$5-10\%,
depending on wavelength, of the total infrared emission (e.g.,
\citealp{treister06}). Here we calculate the total light from
AGN integrated across all wavelengths; 
bolometric corrections to the 2--10~keV luminosity are derived
from the AGN spectral energy
distributions of \citet{treister04}, as described by
\citet{treister06b}. This leads to luminosity-dependent corrections
that range from $\sim$25 for low-luminosity AGN to $\sim$100 for
quasars, in good agreement with observations (e.g.,
\citealp{kuraszkiewicz03,barger05}). Integrating
over the full AGN population, the total bolometric light from AGN is
3.8~nW~m$^{-2}$~sr$^{-1}$. This value is independent of the obscured
AGN fraction, since we are calculating the total AGN output,
regardless of processes of absorption and re-emission at different
wavelengths and depends only on the AGN luminosity function.

Observations of the extragalactic background light integrated over all
wavelengths yielded a value of 55$\pm$20 nW~m$^{-2}$~sr$^{-1}$
\citep{madau00}, while \citet{hauser01} found values in the range
45--170~nW~m$^{-2}$~sr$^{-1}$. Hence, the contribution of AGN to the
total extragalactic light is at most $\sim$8\%. This
contribution can be larger by $\sim$2$\times$ if a large
number of Compton-thick AGN are missed by current X-ray surveys;
however, this is unlikely based on hard X-ray
observations with INTEGRAL and Swift \citep{treister06b,markwardt05}.

According to the Soltan argument \citep{soltan82}, AGN luminosity
traces the accretion of mass onto the central black hole. The
conversion factor between the emitted luminosity and the accreted mass
is the efficiency, $\eta$. Using our estimate of the integrated AGN
luminosity together with the observed local black hole mass density,
$\rho$=4.6$^{+1.9}_{-1.4}~\times$10$^5$~M$_\sun$Mpc$^{-3}$
\citep{marconi04}, we find $\eta$=0.11$^{+0.05}_{-0.03}$. If
instead we use the observed value of
$\rho$=2.9$\pm$0.46~$\times$10$^5$M$_\sun$Mpc$^{-3}$ reported by
\citet{yu02}, we get $\eta$=0.17$\pm$0.02. The derived accretion
efficiency is thus comparable to the typically assumed value,
$\eta\sim$0.1.

In summary, the fraction of obscured AGN increases with redshift,
following an evolution of the form (1+$z$)$^{0.3-0.4}$.  This
evolution could be related to an increase in the dust content of host
galaxies at earlier epochs. The AGN contribution to the total
extragalactic light is small, $\la$8\%. Combining this estimate with
the observed local black hole mass density, we find an average
radiative efficiency of $\eta \sim $0.1-0.2.

\acknowledgments

ET thanks the support of the Centro de Astrof\'{\i}sica FONDAP and
Fundaci\'on Andes. We acknowledge support from NASA grants NNG05GM79G
and HST-GO-09425.13-A.  We thank the referee, Michael Brotherton, for
a useful review of this paper and Megan Eckart for providing us the
SEXSI survey data in electronic format.


\clearpage

\begin{deluxetable}{lcccccccccc}
\tablecolumns{22}
\tablecaption{Constituent Surveys}
\tablehead{
\colhead{Survey} & \colhead{Area} & \colhead{Flux Lim.} & \multicolumn{3}{c}{Sources} & \colhead{Spec.} & \colhead{$m_b$}\tablenotemark{d} & \colhead{$m_l$}\tablenotemark{d} & \colhead{Max.} &\colhead{Ref.}\\
\colhead{} & \colhead{deg$^2$} & \colhead{erg~cm$^{-2}$~s$^{-1}$} & \colhead{Total\tablenotemark{a}} & \colhead{ID\tablenotemark{b}} & \colhead{Obsc.\tablenotemark{c}} & \colhead{Comp.} & \colhead{} & \colhead{} & \colhead{Comp.} & \colhead{} }
\startdata
AMSS & 101.99 & 3$\times$10$^{-13}$ & 79 & 79 & 16 & 100\% & --- & --- & 100\% & 1\\ 
SEXSI & 2.25 & 5$\times$10$^{-14}$ & 1016 & 405 & 194 & 40\% & 20.5 & 26.0 & 68\% & 2\\ 
H2XMM & 0.9 & 2$\times$10$^{-14}$ & 122 & 94 & 33 & 79\% & 23.5 & 28.0 & 100\% & 3\\ 
CLASXS & 0.28 & 4$\times$10$^{-15}$ & 466 & 232 & 129 & 52\% & 21.5 & 27.0 & 100\% & 4\\ 
CYDER & 0.12 & 2$\times$10$^{-15}$ & 124 & 59 & 23 & 50\% & 23.0 & 27.0 & 77\% & 5\\ 
CDF-S & 0.11 & 7$\times$10$^{-16}$ & 231 & 227 & 137 & 99\% & --- & --- & 100\% & 6\\ 
CDF-N & 0.12 & 3$\times$10$^{-16}$ & 303 & 133 & 99 & 66\% & 23.0 & 27.5 & 100\% & 7\\
\enddata
\tablenotetext{a}{Sources selected in the hard X-ray band.}
\tablenotetext{b}{Including only sources with known redshifts and $L_X$$>$10$^{42}$~erg~s$^{1}$.}
\tablenotetext{c}{Sources with no broad lines in the optical spectrum.}
\tablenotetext{d}{Break and limiting magnitudes for a simple fit to spectroscopically identified fraction as a function of optical magnitude.}
\tablerefs{
(1) \citet{akiyama03}; (2) \citet{eckart06}; (3) \citet{fiore03}; (4) \citet{steffen04}; (5) \citet{treister05a}; (6) \citet{zheng04}; (7) \citet{barger03}
}
\end{deluxetable}

\clearpage

\clearpage

\end{document}